\documentclass[12pt]{article}
\usepackage{graphicx}

\begin{document}

\begin{center}
\Large {\bf  Conformal anomalies of CFT's with boundaries}
\end{center}

\bigskip
\bigskip

\begin{center}
D.V. Fursaev
\end{center}

\bigskip
\bigskip

\begin{center}
{\it Dubna State University \\
     Universitetskaya str. 19\\
     141 980, Dubna, Moscow Region, Russia\\

  and\\

  the Bogoliubov Laboratory of Theoretical Physics\\
  Joint Institute for Nuclear Research\\
  Dubna, Russia\\}
 \medskip
\end{center}

\bigskip
\bigskip

\begin{abstract}
The trace anomaly of conformal field theories in four dimensions 
is characterized by '$a$' and '$c$'-functions. The scaling properties  
of the effective action of a CFT in the presence of boundaries is shown to be determined
by $a$, $c$ and two new functions (charges) related to boundary effects. The boundary charges are computed for different theories and different boundary conditions.
One of the boundary charges depends on the bulk $c$ charge.
\end{abstract}

\newpage

\section{Anomaly of the effective action}\label{bc1}

Conformal anomalies are well-known. In a conformal field theory (CFT) in four dimensions the expectation value of the trace of
stress energy tensor can be written
in the following universal form \cite{Duff:1977ay}: 
\begin{equation}\label{1.2}
\langle T_\mu^\mu\rangle=-2a~E- c~ I-{c \over 24\pi^2} ~\nabla^2 R~~,
\end{equation}
\begin{equation}\label{1.3a}
E={1 \over 32\pi^2}\left[R^2-4R_{\mu\nu}R^{\mu\nu}+
R_{\mu\nu\lambda\rho}R^{\mu\nu\lambda\rho}\right]~~,
\end{equation}
\begin{equation}\label{1.3}
I=-{1 \over 16\pi^2}C_{\mu\nu\lambda\rho}C^{\mu\nu\lambda\rho}~~,
\end{equation}
where $E$ is the volume density of the Euler characteristics of the underlying
background manifold $\cal M$, and $C_{\mu\nu\lambda\rho}$ is the Weyl tensor of $\cal M$. 
One can also define the integral anomaly by the variation of the effective 
action $W$,
\begin{equation}\label{1.4a}
{\cal A}\equiv \partial_\sigma W[e^{2\sigma}g_{\mu\nu}]_{\sigma=0}=\int_{\cal M}
\sqrt{g}d^4x \langle T_\mu^\mu\rangle~~,
\end{equation}
under scaling with a constant factor $\sigma$. The right hand side 
of (\ref{1.4a}) relates $\cal A$ to the trace anomaly (\ref{1.2}) and
holds on a closed manifold. 

By analogy with two-dimensional CFT's constants $a$ and $c$ are called charges,
or $a$-function and $c$-function, when they are allowed to 
run under renormalization. There are arguments \cite{Komargodski:2011vj} that $a$-function changes monotonically when going from a critical point to a critical point.

The aim of this work is to study integral anomaly of the effective action  
(\ref{1.4a}) when $\cal M$ has a 
boundary. In this case the right hand side of (\ref{1.4a}) cannot be written solely as the integral of the trace anomaly since new boundary terms may appear.
An experience with two dimensional CFT's
shows that boundary terms in some quantum quantities, such as a $g$-function in the entropy, 
may be interesting from the point of view of the renormalization group \cite{Affleck:1991tk}, \cite{Friedan:2003yc}.

We study different CFT's in four dimensions with the boundary conditions which do not 
break the conformal invariance on the classical level and show that the integral anomaly has the following universal structure:
\begin{equation}\label{1.4}
{\cal A}=-2a~\chi_4-c~i+q_1j_1+q_2j_2~~.
\end{equation}
Quantities $\chi_4$, $i$,
$j_1$ and $j_2$ are scale invariant functionals,
$\chi_4$ is the Euler characteristics of the background manifold, $i$ is the integral of $I$
over $\cal M$, see (\ref{1.3}). The boundaries result 
in two new terms with constants $q_1$ and $q_2$ (the boundary charges). The boundary functionals in (\ref{1.4}) are
\begin{equation}\label{1.10b}
j_1={1 \over 16\pi^2} \int_{\partial {\cal M}}\sqrt{H}d^3x ~C_{\mu\nu\lambda\rho}N^\nu N^\rho\hat{K}^{\mu\lambda}
\equiv {1 \over 16\pi^2} \int_{\partial {\cal M}}\sqrt{H}d^3x G_1~~,
\end{equation}
\begin{equation}\label{1.10a}
j_2={1 \over 16\pi^2} \int_{\partial {\cal M}}\sqrt{H}d^3x~
\mbox{Tr}(\hat{K}^3)\equiv {1 \over 16\pi^2} \int_{\partial {\cal M}}\sqrt{H}d^3x G_2~~.
\end{equation}
We use the following notations: $g_{\mu\nu}$ is the metric of the background manifold $\cal M$,  the metric induced on the boundary $\partial {\cal M}$ of $\cal M$ is
$H_{\mu\nu}=g_{\mu\nu}-N_{\mu}N_\nu$, 
$N^\nu$ is a unit outward pointing normal vector to $\partial {\cal M}$, $K_{\mu\nu}=H_{\mu}^\lambda H_{\nu}^\rho N_{\lambda;\rho}$ is the extrinsic curvature tensor
of $\partial {\cal M}$, 
$\hat{K}_{\mu\nu}=K_{\mu\nu}-H_{\mu\nu}K/3$ is a traceless part of $K_{\mu\nu}$.
Conformal invariance of $j_1$,  $j_2$ follows from the fact
that $\hat{K}_{\mu\nu}$ transforms homogeneously under conformal transformations.
Definitions of the Riemann tensor, the Ricci tensor, the scalar curvature, respectively, are 
$R^\lambda_{\mu\nu\rho}=-\Gamma^\lambda_{\mu\nu,\rho}+...$,  
$R_{\mu\nu}=R^\lambda_{\mu\lambda\nu}$,  $R=R^\mu_\mu$.
Our definitions of $R$ and $K_{\mu\nu}$ coincide with those used by Dowker and Schofield in  \cite{Dowker:1989ue} and by Vassilevich in \cite{Vassilevich:2003xt}. The only difference with those works  
is in the sign of $R^\lambda_{\mu\nu\rho}$ and direction of $N^\mu$. In our paper, in 
\cite{Dowker:1989ue}, and in \cite{Vassilevich:2003xt} the scalar curvature 
on $S^n$ is positive. The curvatures constructed with the metric of $\partial {\cal M}$ will be denoted as $\hat{R}^\lambda_{\mu\nu\rho}$, $\hat{R}_{\mu\nu}$, 
$\hat{R}$.

\begin{table}
\renewcommand{\baselinestretch}{2}
\medskip
\caption{Charges in the anomaly of the effective action}
\bigskip
\begin{centerline}
{\small
\begin{tabular}{|c|c|c|c|c|c|}
\hline
$\mbox{Theory}$  & $a$ & $c$ & $q_1$  & $q_2$ & $\mbox{boundary
condition}$   \\
\hline
$\mbox{real scalar}$ &  ${1 \over 360}$
&  ${1 \over 120}$ & ${1 \over 15}$ 
 &  ${2\over 35}$   & $\mbox{Dirichlet}$ \\
\hline
$\mbox{real scalar}$ &  ${1 \over 360}$
&  ${1 \over 120}$ & ${1 \over 15}$ 
 &  ${2\over 45}$   & $\mbox{Robin}$ \\
\hline
$\mbox{Dirac spinor}$ &  ${11 \over 360}$ 
&  ${1 \over 20}$
 &  ${2 \over 5}$ &
 ${2 \over 7}$ & $\mbox{mixed}$  \\
\hline
$\mbox{gauge Boson}$ &  ${31 \over 180}$
&  ${1 \over 10}$
&  ${12 \over 15}$ & 
${16 \over 35}$ & $\mbox{absolute}$  \\
\hline
$\mbox{gauge Boson}$ &  ${31 \over 180}$
&  ${1 \over 10}$
&  ${12 \over 15}$ & 
${16 \over 35}$ & $\mbox{relative}$  \\
\hline
\end{tabular}}
\bigskip
\renewcommand{\baselinestretch}{1}
\end{centerline}
\label{t1}
\end{table}

Our results for the boundary charges in (\ref{1.4}) are listed in Table \ref{t1}.
$q_2$ can depend on the type of 
the boundary conditions.

It should be noted that (\ref{1.4}) for a scalar field
with the Dirichlet boundary condition follows from results by Dowker and Schofield
\cite{Dowker:1989ue}. Table \ref{t1} agrees with \cite{Dowker:1989ue} for this case.
Anomalous rescalings of the effective action of Dirac fields and gauge bosons
have been also studied by Moss and Poletti \cite{Moss:1994jj} for Einstein spaces with 
boundaries.

To our knowledge the model independent nature of Eq. (\ref{1.4}) has not been emphasized  so far.

In Sec. \ref{hc} we relate the anomaly to known computations of boundary terms
in the corresponding heat kernel coefficients. Anomalies in the models presented in Table \ref{t1}
are discussed in Sec. \ref{bc}. Concluding remarks are given in 
in Sec. \ref{cr}. We point out a universal relation  
between  $c$ and $q_1$ (valid at least for all above models). It hints that $q_1$ is not 
an independent new charge.

\section{The heat coefficient}\label{hc}
\setcounter{equation}0

We use the relation between the anomaly and the heat coefficient 
of a Laplace operator $\Delta=-\nabla^2+X$ for the corresponding conformal theory
\begin{equation}\label{1.4b}
{\cal A}= \eta A_4~~,
\end{equation}
where $\eta=+1$ for Bosons and $\eta=-1$. We ignore in (\ref{1.4b}) a possible contribution
of zero modes of $\Delta$.
The heat coefficients for the asymptotic expansion of the heat kernel of $\Delta$ are defined as
\begin{equation}\label{1.1}
K(\Delta;t)=\mbox{Tr}~e^{-t\Delta}\simeq \sum_{p=0} A_p(\Delta)~t^{(p-4)/2}~~,~~t\to 0~~.
\end{equation}
If the classical theory is scale invariant the heat coefficient $A_4$ is 
a conformal invariant, see e.g. \cite{Fursaev:2011zz}. Therefore $A_4$ can be represented
as a linear combination of conformal invariants constructed of the geometrical characteristics of $\cal M$, $\partial {\cal M}$ and embedding of $\partial {\cal M}$ in
$\cal M$. On dimensional grounds, these invariants (in four dimensions) are $\chi_4$,
$i$, $j_1$ and $j_2$. The bulk part of  $A_4$ is determined by $\chi_4$ and
$i$, and it is well known.  

We are interested in boundary terms in $A_4$.  To find them one 
should take into account the boundary term in $\chi_4$. The definition of
the Euler number for a four-dimensional manifold with a
boundary is as follows:
\begin{equation}\label{1.5}
\chi_4=B_4[{\cal M}]+S_4[\partial {\cal M}]~~,
\end{equation}
\begin{equation}\label{1.6}
B_4[{\cal M}]=\int_{\cal M}\sqrt{g}d^4x~E~~,
\end{equation}
\begin{equation}\label{1.7}
S_4[\partial {\cal M}]={1 \over 32\pi^2}\int_{\partial {\cal M}}\sqrt{H}d^3xQ~~,
\end{equation}
\begin{equation}\label{1.8}
Q=-8\left[\det K_{\mu\nu}+\hat{G}^{\mu\nu}K_{\mu\nu}\right]~~,~~
\hat{G}^{\mu\nu}=\hat{R}^{\mu\nu}-\frac 12H^{\mu\nu} \hat{R}~~.
\end{equation}
Derivation of these formulae can be found in \cite{Dowker:1989ue}. Some values of $\chi_4$ are:
$\chi_4=1$, if $\cal M$ is a domain in $R^3$ with the spherical boundary 
$\partial {\cal M}=S^3$, $\chi_4=0$, if $\cal M$ is a domain 
of a  torus $S^1\times R^3$ with a boundary $\partial {\cal M}=S^1\times S^2$.

The boundary part of $A_4$, therefore, is
\begin{equation}\label{1.9}
A_4^{\mbox{\tiny{bd}}}=\eta(q_1j_1+q_2j_2-2aS_4)~~.
\end{equation}
It is the aim of computations to check that the coefficient at $S_4$ 
does equal $-2\eta a$, where $a$ is the same constant which appears in the trace anomaly
(\ref{1.2}). 

Appearance of $S_4$ among counter terms in a one-loop effective action
dates back to works in 1980's, see \cite{Fradkin:1981iu}.
If ${\cal M}$ is a domain of a flat Euclidean background and $\partial {\cal M}=S^3$
one easily finds for (\ref{1.4}) that ${\cal A}=-2a$. In this case the anomaly is solely 
determined by $S_4$.
Mode-by-mode computations of the anomaly for the spinor and gauge fields 
have been done in \cite{D'Eath:1990td},\cite{Esposito:1994bv}.

Our starting point is  formula (5.33) from Vassilevich's review \cite{Vassilevich:2003xt}
for $A_4$ for mixed boundary conditions. We put there $f=1$, $n=4$.  The boundary conditions are
\begin{equation}\label{1.20}
(\nabla_N-S)\Pi_+\phi=0~~,~~\Pi_-\phi=0~~~,
\end{equation}
where $\nabla_N=N^\mu\nabla_\mu$, $\Pi_\pm$ are corresponding projectors,
$\Pi_++\Pi_-=1$, definition of $S$ coincides with \cite{Vassilevich:2003xt}.

After converting total derivatives in the bulk into surface terms and
some algebra the boundary part $A_4^{\mbox{\tiny{bd}}}$ of $A_4$ can be written as
\begin{equation}\label{1.11}
A_4^{\mbox{\tiny{bd}}}={1 \over (4\pi)^2}\int_{\partial {\cal M}}\sqrt{H}d^3x~
\mbox{Tr}~C_4~~,
\end{equation}
\begin{equation}\label{1.11a}
C_4=\Pi_+C_4^++\Pi_-C_4^-+C_4^{+-}~~,
\end{equation}
$$
C_4^+=-{1 \over 360}Q+{1 \over 15}G_1+{2 \over 45}G_2-\frac 13\left(X-\frac 16 R\right)K+\frac 12\nabla_N\left(X-\frac 16 R\right)+
$$
\begin{equation}\label{1.13}
\frac 43 \left(S\Pi_+ +\frac 13 K\right)^3-2\left(X-\frac 16 R\right)S+
\left(S+\frac 13 K\right)\left({2 \over 15}\mbox{Tr}K^2-
{2 \over 45}K^2\right)~~,
\end{equation}
\begin{equation}\label{1.12}
C_4^-=-{1 \over 360}Q+{1 \over 15}G_1+{2 \over 35}G_2-\frac 13\left(X-\frac 16 R\right)K-\frac 12\nabla_N\left(X-\frac 16 R\right)~~,
\end{equation}
$$
C_4^{+-}=-\frac 13(\Pi_+-\Pi_-)\Pi_{+:a}\Omega_{a\mu}N^\mu-
$$
\begin{equation}\label{1.14}
{2 \over 15}\Pi_{+:a}\Pi_{+:a}K
-{4 \over 15}\Pi_{+:a}\Pi_{+:b}K^{ab}-\frac 43\Pi_{+:a}\Pi_{+:a}\Pi_+S~.
\end{equation}
Here we use 'flat' indices $a,b$ in the tangent space 
to the boundary, see \cite{Vassilevich:2003xt}. $\Omega_{\mu\nu}$ is the field strength of the connection defined by (2.10) 
in \cite{Vassilevich:2003xt}. The notations are 
$\mbox{Tr}K^m=K^{\mu_1}_{\mu_2}...K^{\mu_p}_{\mu_1}$, $K=\mbox{Tr}K$.

In deriving (\ref{1.11a})-(\ref{1.14}) we used Gauss-Codazzi identities and the relations: 
\begin{equation}\label{2.3a}
G_1=R_{\mu\nu\lambda\rho}K^{\mu\lambda}N^\nu N^\rho-\frac 12R_{\mu\nu}(N^\mu N^\nu K+K^{\mu\nu})+\frac 16 KR~~,
\end{equation}
\begin{equation}\label{2.3b}
G_2=\mbox{Tr}K^3-K\mbox{Tr}K^2+\frac 29 K^3~~,
\end{equation}
$$
Q=8R_{\mu\nu\lambda\rho}K^{\mu\lambda}N^\nu N^\rho-8R_{\mu\nu}(N^\mu N^\nu K+K^{\mu\nu})+
4KR+
$$
\begin{equation}\label{2.3}
\frac 83 K^3+{16 \over 3} \mbox{Tr}K^3-8K \mbox{Tr}K^2~~,
\end{equation}
which can be easily
obtained from definitions (\ref{1.10b}),(\ref{1.10a}),(\ref{1.8}).

Representation of the boundary terms in form (\ref{1.11a})-(\ref{1.14})
follows Moss and Poletti \cite{Moss:1994jj}. Calculations of $A_4$ in case of boundaries
have been done by several authors. The key paper is by Branson and Gilkey \cite{Branson}. A complete list 
of references can be found in \cite{Vassilevich:2003xt}.

\section{Computations}\label{bc}
\setcounter{equation}0

\subsection{Conformal scalar field}

In this case $X=1/6$. For the Dirichlet condition $\Pi_+=0$, $C_4^{+-}=0$, and the boundary charges
are determined by $C_4^-$, (\ref{1.12}). 

Conformally invariant Robin condition requires
$S=-K/3$, $\Pi_+=1$. Then again $C_4^{+-}=0$, and boundary charges
follow from (\ref{1.13}).

\subsection{Massless Dirac spinor}

In case of a massless Dirac field $\psi$ the operator is
$\Delta^{(1/2)}=(i\gamma^\mu\nabla_\mu)^2$. The boundary conditions
are mixed ones,
\begin{equation}\label{2.10-dd}
\Pi_- \psi\mid_{\partial {\cal M}}=0~~,~~
(\nabla_N+K/2)\Pi_+ \psi\mid_{\partial {\cal M}}=0~~,
\end{equation}
where $\Pi_\pm=\frac 12 (1\pm i\gamma_\ast N^\mu \gamma_\mu)$, and 
$\gamma_\ast$ is a chirality gamma matrix. Therefore, $X=R/4$, $S=-\Pi_+K/2$. 

The physical meaning of (\ref{2.10-dd})
is that the normal component of the spinor current vanishes on the
boundary. Condition (\ref{2.10-dd}) does not break conformal invariance.
The strength of the spin connection is
$$
\Omega_{\mu\nu}=\frac 14 R_{\mu\nu\sigma\rho}\gamma^\sigma\gamma^\rho~~.
$$
The rest computation is straightforward. One finds
$$
\mbox{Tr}(\Pi_+C_4^++\Pi_-C_4^-)=
$$
\begin{equation}\label{1.21}
r\left(-{1 \over 360}Q+{1 \over 15}G_1+{1 \over 315}G_2
+{1 \over 72}KR+{1 \over 1620} K^3
-{1 \over 90} K\mbox{Tr}(K^2)\right)~~~,
\end{equation}
\begin{equation}\label{1.22}
\mbox{Tr}(C_4^{+-})=r\left(-{1 \over 12}R_{\mu\nu\lambda\rho}K^{\mu\lambda}N^\nu N^\rho
-{1 \over 15} \mbox{Tr}(K^3)
+{1 \over 20} K\mbox{Tr}(K^2)\right)~~,
\end{equation}
where $r=4$ is the number of components of the Dirac spinor in four dimensions.
In computing $C_4^{+-}$ one uses the relation 
$$
\Pi_{+:a}=\frac i2\gamma^b\gamma_\ast K_{ba}~~.
$$
The data of Table \ref{t1} follow from the sum of (\ref{1.21}) and (\ref{1.22}) 
and relations (\ref{2.3a})-(\ref{2.3}).

\subsection{Gauge boson}

By following \cite{Vassilevich:2003xt} we consider the quantization of an
Abelian gauge field $V_\mu$ in the Lorentz gauge 
$\nabla V=0$. The results of Table \ref{t1} are valid for a gauge invariant
combination 
\begin{equation}\label{1.4b}
{\cal A}=  A_4(\Delta^{(1)})-2A_4(\Delta^{(\mbox{\tiny gh})})~~,
\end{equation}
where $(\Delta^{(1)})^\nu_\mu=-\nabla^2\delta^\nu_\mu+R^\nu_\mu$ is the vector Laplacian,
and $\Delta^{(\mbox{\tiny gh})}=-\nabla^2$ is the 
Laplacian for ghosts.

We study two sorts of boundary conditions. The absolute  (or electric
\cite{Moss:1994jj}) boundary condition:
\begin{equation}\label{2.10-el}
N^\mu F_{\mu\nu}\mid_{\partial {\cal M}}=0~~,
\end{equation}
and relative (or magnetic \cite{Moss:1994jj}) boundary condition:
\begin{equation}\label{2.10-ma}
N^\mu \tilde{F}_{\mu\nu}\mid_{\partial {\cal M}}=0~~,
\end{equation}
where $F_{\mu\nu}=\nabla_\mu V_\nu-\nabla_\nu V_\mu$, and $\tilde{F}_{\mu\nu}$ is the Hodge
dual of $F_{\mu\nu}$.

The both conditions are manifestly gauge and conformally 
invariant. 

Condition (\ref{2.10-el}) requires that components of an electric field normal to $\partial {\cal M}$
and components of the magnetic field which are tangential to $\partial {\cal M}$
vanish on the boundary. This means that the boundary is a perfect 
conductor. In condition (\ref{2.10-ma}) the roles of electric and magnetic fields are interchanged.

In the Lorentz gauge the absolute boundary condition is reduced to
the following conditions on the vector field and ghosts \cite{Vassilevich:2003xt}: 
\begin{equation}\label{1.23}
(\delta_\mu^\nu \nabla_N
+K_{\mu}^\nu)V_\nu^{+}\mid_{\partial {\cal M}}=0~~,~~
V^{-}_\mu\mid_{\partial {\cal M}}=0~~,
\end{equation}
where $V^{\pm}=\Pi_\pm V$, and
\begin{equation}\label{1.24}
(\Pi_+)_\mu^\nu=\delta_\mu^\nu-N_\mu N^\nu~~,~~(\Pi_-)_\mu^\nu=N_\mu N^\nu~~.
\end{equation}
The corresponding boundary condition for a ghost field $c$ is
\begin{equation}\label{1.25}
\partial_N c\mid_{\partial {\cal M}}=0~~.
\end{equation}
The relative condition in the same gauge is
\begin{equation}\label{1.26}
(\nabla_N
+K)V_\mu^{+}\mid_{\partial {\cal M}}=0~~,~~V^{-}_\mu\mid_{\partial {\cal M}}=0~~,~~
\end{equation}
\begin{equation}\label{1.27}
(\Pi_+)_\mu^\nu=N_\mu N^\nu~~,~~(\Pi_-)_\mu^\nu=\delta_\mu^\nu-N_\mu N^\nu~~.
\end{equation}
\begin{equation}\label{1.28}
c\mid_{\partial {\cal M}}=0~~.
\end{equation}

By taking into account that $X^\mu_\nu=R^\mu_\nu$,  $(\Omega_{\lambda\rho})_\mu^\nu
=R^\nu_{\mu\lambda\rho}$, for the vector component with the absolute condition
(\ref{1.23}), (\ref{1.24}) ($S^\mu_\nu=-K^\mu_\nu$) we find
$$
\mbox{Tr}(\Pi_+C_4^++\Pi_-C_4^-)=
$$
\begin{equation}\label{1.29}
(N^\mu R_{\mu a})^{:a}-\frac 16 R_{,N}-{4 \over 360}Q+{4 \over 15}G_1-{8 \over 7}G_2
-{4 \over 9}KR+R_{\mu\nu}(N^\mu N^\nu K+K^{\mu\nu})~~,
\end{equation}
\begin{equation}\label{1.30}
\mbox{Tr}(C_4^{+-})={2 \over 3}R_{\mu\nu\lambda\rho}K^{\mu\lambda}N^\nu N^\rho
+{4 \over 5} \mbox{Tr}(K^3)
-{4 \over 15} K\mbox{Tr}(K^2)~~.
\end{equation}
For the ghost part with condition (\ref{1.25}) $S=0$ and
\begin{equation}\label{1.31}
C_4=C_4^+=-{1 \over 12} R_{,N}-{1 \over 360}Q+{1 \over 15}G_1+{2 \over 45}G_2
+{1 \over 18}KR+{14 \over 405}K^3+{2 \over 45} K\mbox{Tr}(K^2)~~,
\end{equation}
For the vector component with relative conditions (\ref{1.26}), (\ref{1.27}) 
($S^\mu_\nu=-K \delta^\mu_\nu$) 
$$
\mbox{Tr}(\Pi_+C_4^++\Pi_-C_4^-)=-(N^\mu R_{\mu a})^{:a}+\frac 16 R_{,N}
$$
\begin{equation}\label{1.32}
-{4 \over 360}Q+{4 \over 15}G_1+{8 \over 45}G_2
-{4 \over 9}KR+R_{\mu\nu}(N^\mu N^\nu K+K^{\mu\nu})
-{34 \over 405}K^3-{4 \over 45} K\mbox{Tr}(K^2)~~,
\end{equation}
\begin{equation}\label{1.33}
\mbox{Tr}(C_4^{+-})=-{2 \over 3}R_{\mu\nu\lambda\rho}K^{\mu\lambda}N^\nu N^\rho
-{8 \over 15} \mbox{Tr}(K^3)
+{16 \over 15} K\mbox{Tr}(K^2)~~.
\end{equation}
For the ghost part with condition (\ref{1.28}) $S=0$ and
\begin{equation}\label{1.34}
C_4=C_4^-={1 \over 12} R_{,N}-{1 \over 360}Q+{1 \over 15}G_1+{2 \over 35}G_2
+{1 \over 18}KR~~.
\end{equation}
In deriving (\ref{1.29}), (\ref{1.32}) we used the identity
$$
\frac 12 R_{,N}=(N^\mu R_{\mu a})^{:a}+KR_{\mu\nu} N^{\mu}N^{\nu}-R_{\mu\nu} K^{\mu\nu} + N^\mu N^\nu N^\lambda R_{\mu\nu;\lambda}~~.
$$
The first term in the right hand side of this identity is the total derivative 
on $\partial {\cal M}$. Let us emphasize that the two sets of boundary conditions for the gauge field result in the same boundary terms in the anomalous scaling of the effective action.

\section{Concluding remarks}\label{cr}
\setcounter{equation}0

The aim of this paper was to demonstrate a model independent form of the integral
anomaly (\ref{1.4}) in the presence of boundaries and obtain specific values
of the boundary charges for some typical CFT's. It would be important now to study 
evolution of the boundary charges under the renormalization group. 

Computations of boundary charges for other models can be continued 
along the lines of the present paper. Extensions of (\ref{1.4}) to higher dimensional CFT's, say to 6 dimensions, are possible. Since the number of scale invariant structures
is increasing we expect more boundary charges. A principal challenge here is 
the knowledge of boundary terms in the heat coefficient $A_6$.

An interesting issue is a possible relation between bulk charges $a$, $c$ and boundary 
charges $q_1$, $q_2$. A conjecture of \cite{Solo-2015} is that $q_1$ and $c$ may be related. Indeed, all models presented in Table \ref{t1} satisfy a universal 
relation
\begin{equation}\label{conj}
q_1=8c~~.
\end{equation}
We leave a general proof and implications of (\ref{conj}) for a future analysis.

It should be noted that formula (\ref{1.4}) was also discussed in a recent work \cite{Herzog:2015ioa} which appeared several days earlier of the present publication. 

\bigskip
\bigskip
\bigskip

\noindent
{\bf Acknowledgement}

\bigskip

The author is grateful to I.L. Buchbinder and D.V. Vassilevich  
for helpful discussions, A.A. Tseytlin, G. Esposito and K. Jensen for useful comments.
Many thanks go to S.N. Solodukhin for a remark on the connection of $q_1$ with $c$. 
The remark also
helped to fix misprints in Table \ref{t1} for the spinor field.
This work was supported by RFBR grant 13-02-00950.

\end{document}